\documentclass[journal=jacsat,manuscript=article]{achemso}
\usepackage[version=3]{mhchem}
\usepackage{graphicx}

\author{Ajith R,\, Vincent Mathew}
\affiliation[Central University of Kerala]
{Department of Physics, Central University of Kerala, 
 Kasaragod,   
Kerala - 671 314, INDIA}
\author{P.Arun}
\affiliation[S.G.T.B. Khalsa College, University of Delhi]
{Mater. Sci. Res. Lab., S.G.T.B. Khalsa College, University of Delhi, 
Delhi 110007, INDIA}
\email{arunp92@physics.du.ac.in}

\title[\texttt{achemso} demonstration]
{Surface Plasmon Resonance of Dumb-bell Nano-structure}

\begin{document}
\begin{abstract}
We present an intuitive theoretical description of the optical
properties of complex metal nano-structure, consisting of two nano-shells 
connected by a nano-rod giving a dumb-bell like
appearance. The simulations were done using Finite Element Method. The effect 
of the nano-rod length and radius as also the dimensions of the nano-shells
were analyzed. The absorption spectra as in peak positions and intensities
have been found to have a strong dependence on the geometrical parameters of 
the dumb-bell. This study provides evidence that the localized surface 
plasmon modes play a key role in the broadband light harvesting capabilities 
of these nanostructures and is promising for a wide range of practical 
applications, for example in surface-enhanced spectroscopies.
\end{abstract}

\section{Introduction}

Surface plasmons are electromagnetic surface waves originating due to the 
collective oscillation of the conduction electrons near the metal surface
\cite{maier2007plasmonics,raether1988surface}. Light coupling with surface 
plasmons results in enhancement of the local  electromagnetic field and 
strong resonance in the extinction profile of conductive nano-particles 
leading to the optical phenomenon known as localized surface plasmon 
resonance (LSPR) \cite{brongersma2007surface,sarid2010modern}. This 
interaction produces coherent localized plasmon oscillations with a resonant 
frequency that strongly depends on the  shape and size, as also the 
dielectric media surrounding it \cite{maier:011101}. While the effect of size 
and influence of dielectric media have been studied quite extensively, both 
experimentally \cite{0957-4484-17-14-023} and theoretically 
\cite{doi:10.1021/nl8021835}, this has not been the case when it comes to
experimentally studying the effect of shape. Most of the nano 
particles usually obtained are spherical in nature or are rod like with 
circular cross-section. Theoretically, both shapes are interesting due to 
their symmetry. While the position of SPR peaks obtained from spherical grain 
is controllable by controlling the radius of sphere, the nano-rods present  
two variables, rod-length and radius of cross-section. Thus, one expects
higher degree of control with more complex shaped metal clusters.
One rarely comes across complex shaped metal clusters experimentally, thus 
making inroads along these lines difficult. However, with maturing of the 
methods used to theoretically simulate the
optical properties of metal nano-clusters \cite{0034-4885-70-1-R01} exotic 
shapes have been reported, for example, nanorods with triangular cross-section 
\cite{doi:10.1021/nl9036627,doi:10.1021/nl0156888}, pyramid 
shaped \cite{doi:10.1021/ar800126p} grains etc.

While, nano-particles embedded in dielectric media with fairly constant
dielectric constant over frequency ranges of interest are well studied, the 
case where it varies with frequency has not attracted much attention.
However, situations in which a nanoparticle having another nanoparticle in 
its vicinity whose dielectric constant varies with frequency can lead to 
interesting LSPR features \cite{jain}. For example the two nanoparticles 
would interact electromagnetically with each other to show high field 
enhancement in the gap. Recently, a bow tie nanoparticle, formed from  
equilateral triangle shaped nano particles, 
excited at its LSPR wavelength has been reported to generate extremely large 
fields within the gap \cite{Li20122223,sarid2010modern}. 
This has been exploited to study Raman spectra for samples which would usually 
give weak signals in Surface-enhanced Raman Spectroscopy (SERS). This
clearly shows studying exotic structures can also lead to applications of 
importance. 

Hence, in this manuscript we undertake to further our understanding of 
dumb-bell shaped nano-particles that were experimentally reported recently.
Wang et al
\cite{Wang} and Kuldeep et al \cite{kapil} have reported experimental evidence
of SPR in dumb-bell shaped clusters formed from spherical nano particles. 
While Wang et al's clusters were of gold, Kuldeep et al reported two 
core-shell (core of Cesium Bromide and the shell of metal Cesium) connected 
by nano-rod bridges of Cesium. 
Though Kuldeep et al did not have control over the size of the 
dumb-bells (as compared to Wang et al), their results were interesting since 
the exotic shape allowed for four possible control parameters,
namely, the length (l) of the nano-rod bridge separating the two spherical
core-shell structure, the cross-sectional radius (r) of the nano-rod, the
dielectric spherical core's radius (${\rm R_i}$) and  the core-shell's 
radius (${\rm R_o}$), see fig~1. The shell thickness would be given by (${\rm
R_o-R_i}$).

In this paper, the localized surface plasmon resonance of a dumb-bell structure 
is investigated using Finite Element Method. The dependence of plasmon 
resonance on various geometrical parameters are discussed. For field 
polarization orthogonal and parallel to the interparticle symmetry axis, the 
spectral properties of the structure are usually different. In this paper, 
the behavior of the dumbbell structure in presence of an unpolarized incident 
electromagnetic radiation is studied analogous to the usual experimental 
conditions. Nevertheless, studies with polarized light are included whenever 
required. The next section explains the simulation model used in the 
calculations. The section ``Results and Discussion'' provides the detailed 
analysis of the behavior of surface plasmon resonance on
various geometrical parameters.

\section{Simulation}
The simulations were done using the commercially available Finite Element
Method (FEM) package COMSOL Multiphysics $4.2$ (with the RF module) along 
with Matlab $(R2010a)$. The three dimensional simulation domain was composed 
of four spherical volumes: a core, a shell, an embedding medium, and a 
perfectly matched layer (PML) in addition to a cylinder connecting the 
shells as shown in fig~1. The nanoparticle core was taken to be $CsBr$, 
enclosed in a $Cs$ shell modeled using the empirically determined bulk 
dielectric constants provided by Palik \cite{palik1985handbook} with linear 
interpolation. The medium surrounding the nanosystem  was considered to be 
air. A plane wave was used for excitation of nanostructure and was inserted 
on the inside of PMLs surrounding the embedding medium. The dimensions of the 
embedding volume and 
the PML were chosen such that increase in the dimension of the
nanoparticle further would not affect the simulation results.
Discretization of the simulation domain was performed using the built-in
meshing algorithm in COMSOL, which partitioned the simulation space into a 
collection of tetrahedral finite elements.
Large field enhancements possible due to plasmon resonances, required the
application of mesh parameters to ensure convergence of the simulation in the 
required frequency regime.  

\section{Results and Discussion}

In the following passages we discuss the variation in the absorption spectra
of a single dumb-bell shaped CsBr-Cs cluster. Since the 
absorption spectra of the structure depends on four variables, namely l, r, 
${\rm R_i}$ and ${\rm R_o}$, we have stimulated the spectra keeping three 
parameters fixed and one variable at a time.   

\subsection{Effect of Nanorod on Absorption}
Before proceeding to the dumbbell structure, it is worthwhile to investigate 
the surface plasmon resonance peak due to the nano-rod. Under usual
circumstances the nano-rod's length is far greater than its radius, 
${\rm l>>r}$, then due to the anisotropy in its shape, we expect it to have 
two modes of oscillations, namely the transverse mode along the short-axis 
and the longitudinal mode along the long-axis. More explicitly,
when light is incident along the `Y'-axis, as in our case (fig~1), the linear 
polarised light with electric field along `X'-axis (${\rm E_x}$) gives the 
longitudinal mode excitation while that along `Z'-axis (${\rm E_z}$) would 
result in transverse mode excitation. Such polarization dependency has been
reported.\cite{34,41} However, the simulation for a 
Cesium nano-rod having r and l as 10~nm and 40~nm respectively shows just
one broad peak around 570~nm with no evidence of peaks in UV or IR region.
To investigate why we get a broad peak, we have simulated SPR due to the
transverse mode and longitudinal mode separately and find peaks at 500~nm
and 600~nm due to the two modes respectively. The two peaks are close to
each other because of the small dimensions of the nano-rod 
\cite{doi:10.1021/jp9917648}. Also, the longitudinal LSPR is also much 
stronger compared to the transverse 
mode due to the larger polarizability of the nanorod along the longitudinal 
direction. Hence, the aggregate of the two peaks give a single broad peak 
around 570~nm, i.e. tending towards the longitudinal mode's peak position.

To further our understanding on the contribution from the nano-rod, we have
investigated the influence of its size. An interesting observation made was 
that a redshift occurs in the SPR peak for increasing aspect ratio for the 
longitudinal mode while for the transverse mode excitation the peak blueshifts. 
For the unpolarized wave, the peak followed the trend shown by the 
longitudinal mode (fig~2b). As the Transmission Electron Microscope (TEM) 
images of Kuldeep et al\cite{kapil} suggest, their nano-rods were parallel to 
the substrate (`XZ' plane) and the absorption spectra were obtained in 
tranmission mode, implying for unpolarized light both transverse and
longitudinal modes exist. The main results of
their work was that as the as aspect ratio increases the SPR peaks shifted
to lower wavelengths. Which would imply that the transverse mode played a
dominant role there.

We now proceed to investigate the role of nano-rod bridges in the dumb-bell 
structure. For this, we have maintained the outer radius of the shells at 
$50~$nm, inner radius at $20~$nm while varying the length and radius of the 
nano-rod. The curves of fig~3 are distinctly different from that shown
for a nano-rod without shells on its two sides (fig~2a). Not only has an
additional peak appeared in the UV-region, the contributions from the two
structures (shells and nano-rod) seem to interact constructively giving 
resonant absorption peaks whose intensities have increased (comparison shows 
${\rm \times ~50}$ increase). Such intensity/ field enhancements have been
reported \cite{36} with Nie et al showing ${\rm \times~14}$ enhancements due
to constructive coupling of SPRs.\cite{37}

Figure~4 shows the field pattern corresponding to the peaks at UV (fig~4a) and 
visible (fig~4b) range of the spectra for the transverse modes. As explained 
in above passages, the visible peak is due to the plasmon 
resonance along the nanorod, which is enhanced by the two $Cs-CsBr$ spherical 
shells, however, notice that the fields are not concentrated only on the 
cylinder. Infact a coupling of the localized plasmons of the shells 
along the cylinder is also observed possibly leading to the enhancement in 
absorption (fig~4b). Fig~4a show that the UV peak is essentially concentrated 
on the $CsBr-Cs$ interface. The role of the shell's thickness and details
would be discussed in the next section.

Compiling the results, we have shown the variation of absorption peak 
intensity and position for both the UV and visible peak with the nano-rod's 
aspect ratio. As expected, the peak position (${\rm \lambda_{max}}$) of the 
SPR's visible peak shows a linear relation with aspect ratio (Fig~5). Such
size dependent shifts (red-shift with increasing grain size) have been well
documented.\cite{anynano} However, since the shell dimensions were kept
fixed in these simulations, we do not observe any variation in the UV peak's
maxima position.

Fig~6 shows that there is a linear relationship between the 
absorption intensity of two peaks with the aspect ratio. The intensity is due 
to the amount of oscillating electrons present and hence related to the metal 
present in the cluster. As the aspect ratio of the nano-rod increases, metal 
content increases and hence an increase in the absorption intensity. 
Interestingly, these simulations were done for fixed shell dimensions and yet 
the contribution of UV absorption by these spheres increases with increasing 
aspect ratio of the nano-rods. However, one has to remember that the 
metallic nano-rod provides a conductive path. The existence of conductive
path allows for charge transfer plasmons (CTP)\cite{nl1} to the charges 
distributed on the nano-shells. These charges appear due to charge 
accumulation from the longitudinal excitation mode. The large accumulation of 
charges of opposite signs on either sides of the nano-rod gives an enhancement 
of local fields (fig 4~b).\cite{nl2,nl3} Also, the CTP is sensitive to the number of 
opposite signed charges present on either side of the nano-rod and hence on 
the nano-rod's conductivity. Increasing nano-rod length decreases conductive 
path, leading to charge accumulation and hence enhanced absorption intensity.

\subsection{Effect of Core-Shell}

The previous section clearly showed that the shells act as amplifiers,
increasing the plasmon oscillations and hence resonant absorption in the
visible region caused by the nanorods. The shell also contributes a peak in 
the UV region whose absorption intensity is influenced by the aspect ratio of 
nano rod. This section will look into the effect of varying shell size on
the absorption patterns. The simulations in this section were done keeping the 
nano-rod dimensions fixed at $l=r=10~$nm and varying ${\rm R_o}$ and ${\rm
R_i}$, the core-shell dimensions.
Fig~(7) shows some of the simulated absorption spectra, where we have varied 
the outer radius of shells (${\rm R_o}$), keeping the radius of core 
(${\rm R_i}$) constant ($=20~$nm). Alongside this, the compilation of our 
simulations shows the variation of peak position with the shell thickness. 
It can be seen that ${\rm \lambda_{max}}$ of the visible peak exhibits no 
shift with change in shell thickness. This shows that the visible SPR's peak 
position depends solely on the variation of the nano-rod's dimension.
However, the UV peaks shows a linear relation with ${\rm R_o-R_i}$ with the 
slope reflecting a red shift with increasing ${\rm R_o-R_i}$.

Fig~(8) shows that there is a linear relation between the  absorption 
intensity of the peak at UV range and the thickness of the metallic Cesium 
shell, (${\rm R_o-R_i}$). As explained above, an accumulation of
opposite charges takes place on either side of the conductive nano-rod due
to the longitudinal mode of excitation. Thicker shells for a given nano-rod
length mean larger accumulation of charges and hence more absorption as is
reflected by increased absorption peak intensities. 
Interestingly, as the shell size increases (beyond $50~$nm), a SPR peak appears 
in the IR region. Oldenburg et al \cite{halas} have done extensive work on
metal shells and have shown that each surface gives its own plasmon mode.
The two modes couple across the shell thickness with the coupling strength
based on the dipole model follow ${\rm (R_o-R_i)^{-3}}$ trend. Hence, with
increase in shell thickness the coupling between the two surfaces, CsBr-Cs and
Cs-air, would diminish. We believe this diminished coupling gives rise to
the IR peak. We have investigated which surface gives the IR peak using the
field pattern. The field pattern for the peak at the IR region is shown in 
fig~8 for different shell thickness. It can be seen that the IR peak 
corresponding to the plasmon resonance occur at the outer surface of the 
shell. It has been observed previously that the plasmon resonance peak of the 
$Cs$ sphere occurs in the far IR range.\cite{exp} The above analysis
shows that dumb-bell metal nano-structures are highly tunable and suitable
for experimentalist who might require controllable geometrical parameters in
developing light harvesting sensors etc.

\section{Conclusion}
We have investigated the localized surface plasmon resonance of a dumb-bell 
structure consisting of two $CsBr-Cs$ core-shells connected with a $Cs$ 
nano-rod. As a primary step, the plasmon resonance of a $Cs$ nano-rod
is studied and obtained that the plasmon absorption peak resides in the 
visible region. The study of the dumb-bell structure revealed that the two 
nano-shells enhances the absorption in the visible region in addition to
contributing a peak in the UV range. As the shell thickness of the 
dumb-bell structure increased, another peak appeared in the IR region. 
Tunability of these resonance conditions with various geometrical parameters 
were appreciated.

\section*{Acknowledgement}
We would like to acknowledge our gratitude to the University Grants
Commission (UGC, Delhi) for its finanical assistance (F.No. 39-531/2010 SR)
for carrying out this work. Also, the finanical travel grant given by
University of Delhi (Innovation Projects, SGTB-101) is gratefully
acknowledged.

%\subsection{References}

%\bibliography{achemso}

\newpage
\section*{Figure Captions}
\begin{itemize}
\item[1.] Geometry of the dumb-bell structure. Light when incident along 
the `Y'-axis show excitation of the transverse mode due to the electric
field component, ${\rm E_z}$, while ${\rm E_x}$ contributes to excitation of 
the longitudinal mode.
\item[2.] a) Variation of absorption with wavelength for a nanorod, and b)
the variation in LSPR's ${\rm \lambda_{max}}$ with aspect ratio of the 
nanorod.
\item[3.] As an example we show simulated spectra of absorption with
wavelength for different values of length of the rod connecting the shells.
\item[4.] Field distribution in the $XZ$ plane for the structure for
a) UV and b) visible peak.
\item[5.] Figures shows the variation of 
the UV and visible peak's position with metal Cesium
nano-rod's aspect ratio (l/r).
\item[6.] Figure shows the variation of the UV and visible absorption
peak's intensity with nano-rod's aspect ratio. Alongside the field
distribution in the XZ plane is shown for two different nano-rod lengths.
\item[7.] The simulated absorption spectra for increasing Cesium shell
thickness keeping aspect ratio/ length of the nano-rod constant. Compiling
the results of UV and visible peak positions (${\rm \lambda_{max}}$) of the 
simulated spectra show no variation in visible peak position while UV's peak
position shows a steady red-shift.
\item[8.] Figure shows the variation of the UV absorption peak's intensity 
and the field distribution in the $XZ$ plane for the structure for    
outer radius of $50$~nm and $80$~nm.
\end{itemize}

\newpage
%%%%%%%%%%%%%%%%%%%%%%%%%%%%%%%%%%%%%%%%%%%%%%%%%%%%%%%%%%%%%%%%%%%%%%%%%
\begin{figure}[h]
%\vskip -2cm
\begin{center}   
\includegraphics[width=4.75in, angle=-0]{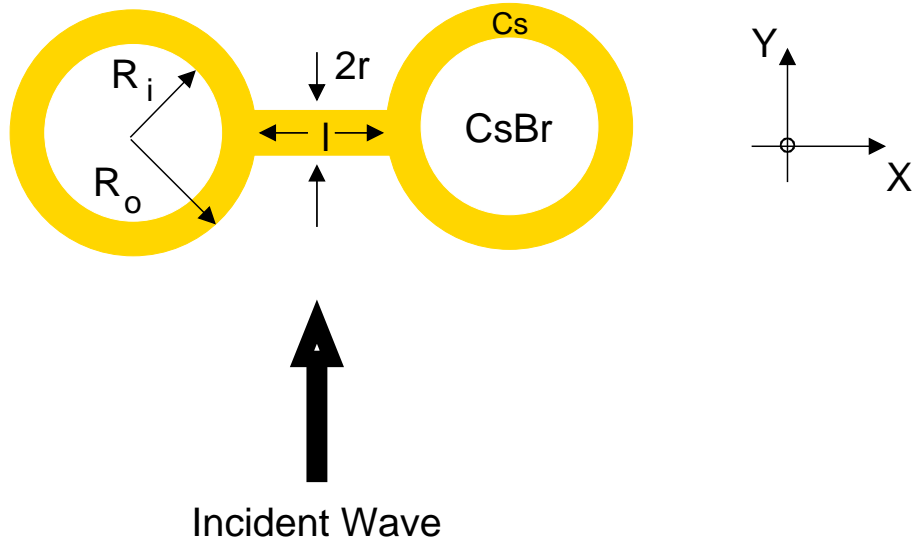}
%\vskip -2.2cm
\caption{\sl Geometry of the dumb-bell structure. Light when incident along 
the `Y'-axis show excitation of the transverse mode due to the electric
field component, ${\rm E_z}$, while ${\rm E_x}$ contributes to excitation of 
the longitudinal mode. }
\end{center}   
\label{1}  
\vskip -1cm
\end{figure}   
%%%%%%%%%%%%%%%%%%%%%%%%%%%%%%%%%%%%%%%%%
\newpage
%%%%%%%%%%%%%%%%%%%%%%%%%%%%%%%%%%%%%%%%%%%%%%%%%%%%%%%%%%%%%%%%%%%%%%%%%
\begin{figure}[h!!]
\begin{center}   
\includegraphics[width=3.15in, angle=-0]{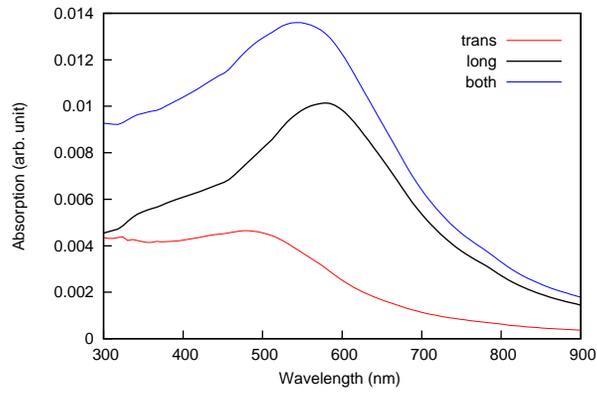}
\vfil
\includegraphics[width=2.15in, angle=-90]{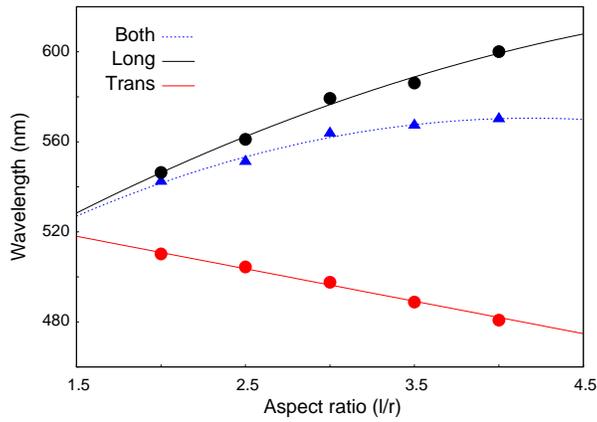}
\caption{\sl a) Variation of absorption with wavelength for a nanorod, and b)
the variation in LSPR's ${\rm \lambda_{max}}$ with aspect ratio of the 
nanorod.}
\end{center}   
\label{cylnew}  
\end{figure}   
%%%%%%%%%%%%%%%%%%%%%%%%%%%%%%%%%%%%%%%%%
\newpage
%%%%%%%%%%%%%%%%%%%%%%%%%%%%%%%%%%%%%%%%%%%%%%%%%%%%%%%%%%%%%%%%%%%%%%%%%
\begin{figure}[h]
\begin{center}   
\includegraphics[width=4in, angle=-0]{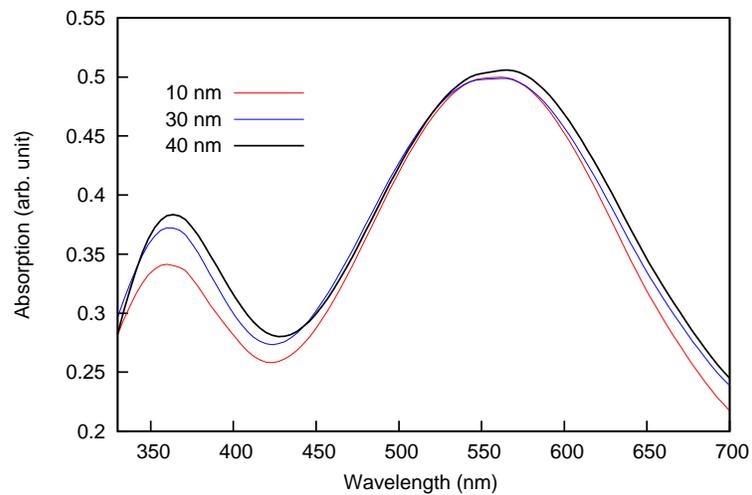}
\caption{\sl As an example we show simulated spectra of absorption with
wavelength for different values of length of the rod connecting the shells.}
\end{center}   
\label{lenrod}  
\end{figure}   
%%%%%%%%%%%%%%%%%%%%%%%%%%%%%%%%%%%%%%%%%
\newpage
%%%%%%%%%%%%%%%%%%%%%%%%%%%%%%%%%%%%%%%%%%%%%%%%%%%%%%%%%%%%%%%%%%%%%%%%%
\begin{figure}[ht!!]
\begin{center}   
%\vskip 1cm
%\includegraphics[width=2.75in, angle=-0]{5020.eps}
%\hfil
\includegraphics[width=6.75in, angle=-0]{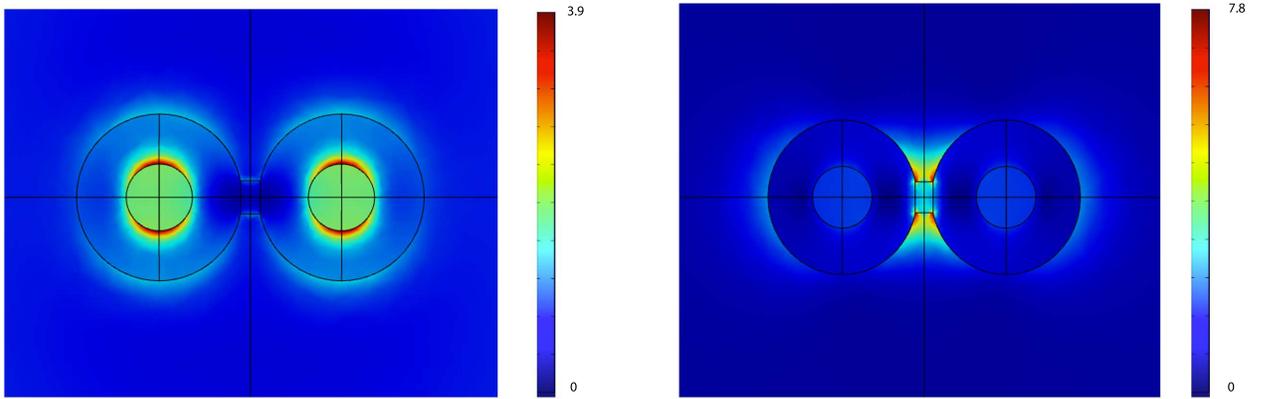}
\vskip 1.6cm
\caption{\sl Field distribution in the $XZ$ plane for the structure for
a) UV and b) visible peak.}
%\vskip -4cm
\end{center}   
\label{rlfixed1}  
\end{figure}   
%%%%%%%%%%%%%%%%%%%%%%%%%%%%%%%%%%%%%%%%%
\newpage
%%%%%%%%%%%%%%%%%%%%%%%%%%%%%%%%%%%%%%%%%%%%%%%%%%%%%%%%%%%%%%%%%%%%%%%%%
\begin{figure}[h]
\begin{center}   
\includegraphics[width=3.25in, angle=-90]{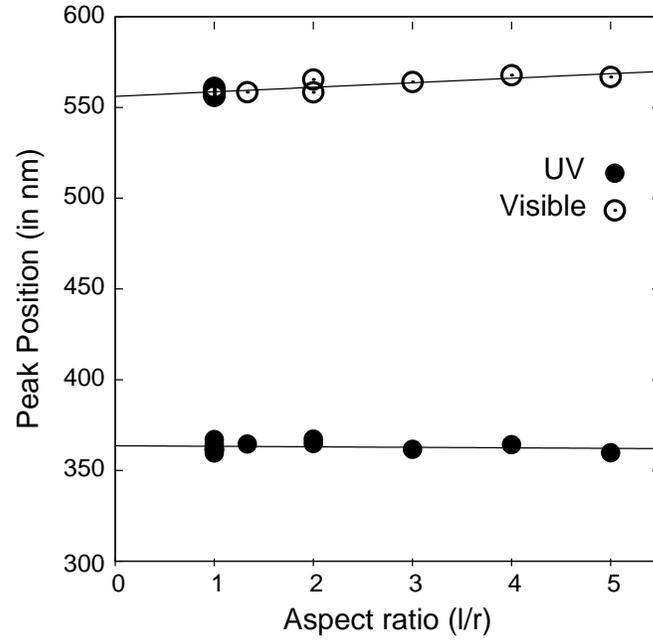}
\caption{\sl Figures shows the variation of 
the UV and visible peak's position with metal Cesium
nano-rod's aspect ratio (l/r).}
\end{center}   
\label{rlfixed11}  
\end{figure}   
%%%%%%%%%%%%%%%%%%%%%%%%%%%%%%%%%%%%%%%%%
\newpage
%%%%%%%%%%%%%%%%%%%%%%%%%%%%%%%%%%%%%%%%%%%%%%%%%%%%%%%%%%%%%%%%%%%%%%%%%
\begin{figure}[h!!]
\begin{center}   
\includegraphics[width=2.75in, angle=-90]{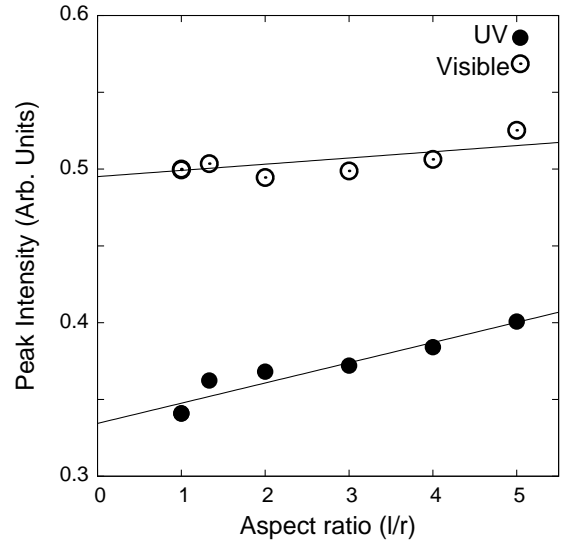}
\vskip 1.6cm
\includegraphics[width=6.75in, angle=-0]{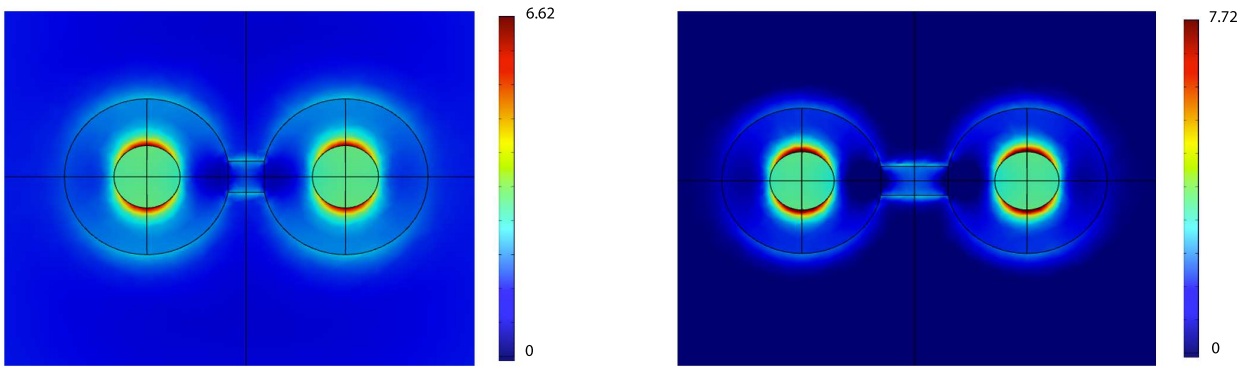}
\vskip 1.7cm
\caption{\sl Figure shows the variation of the UV and visible absorption
peak's intensity with nano-rod's aspect ratio. Alongside the field
distribution in the XZ plane is shown for two different nano-rod lengths. }
\end{center}   
\label{rlfixed1}  
\end{figure}   
%%%%%%%%%%%%%%%%%%%%%%%%%%%%%%%%%%%%%%%%%
\newpage
%%%%%%%%%%%%%%%%%%%%%%%%%%%%%%%%%%%%%%%%%%%%%%%%%%%%%%%%%%%%%%%%%%%%%%%%%
\begin{figure}[h!]
\begin{center}   
\includegraphics[width=3.25in, angle=-0]{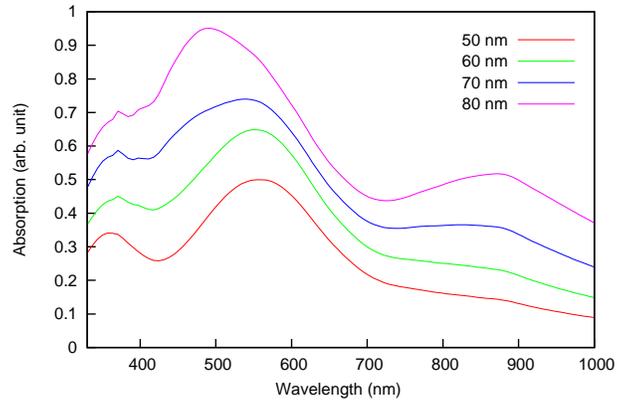}
\vfil
\includegraphics[width=3.25in, angle=-90]{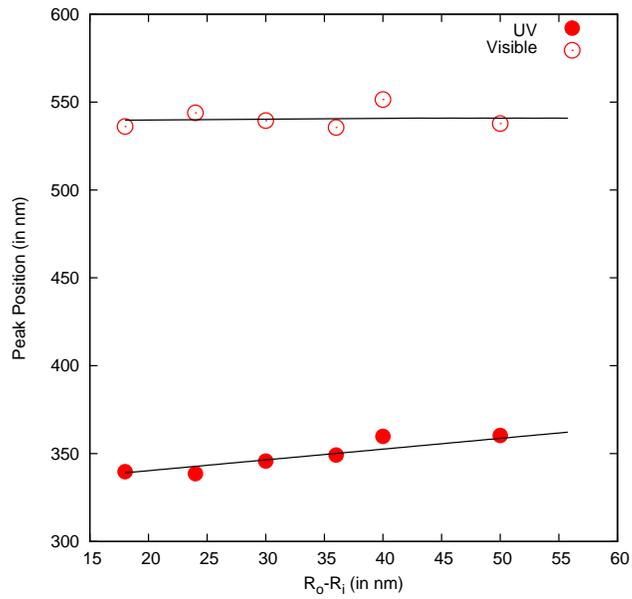}
\caption{\sl The simulated absorption spectra for increasing Cesium shell
thickness keeping aspect ratio/ length of the nano-rod constant. Compiling
the results of UV and visible peak positions (${\rm \lambda_{max}}$) of the 
simulated spectra show no variation in visible peak position while UV's peak
position shows a steady red-shift.}
\end{center}   
\label{rlf}  
\end{figure}   
%%%%%%%%%%%%%%%%%%%%%%%%%%%%%%%%%%%%%%%%%
\newpage
%%%%%%%%%%%%%%%%%%%%%%%%%%%%%%%%%%%%%%%%%%%%%%%%%%%%%%%%%%%%%%%%%%%%%%%%%
\begin{figure}[h!]
\begin{center}   
\includegraphics[width=2.75in, angle=-90]{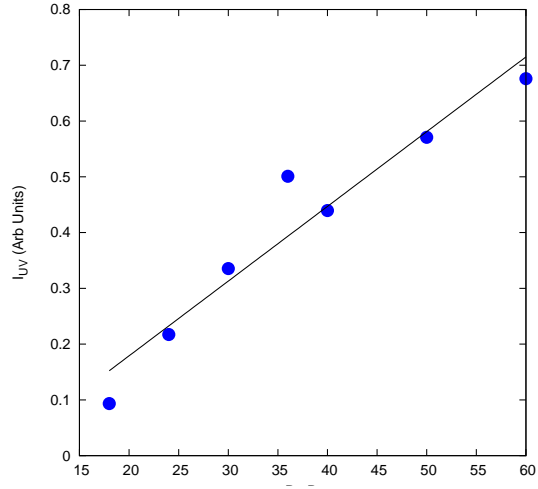}
\vskip 1.6cm
\includegraphics[width=6.75in, angle=-0]{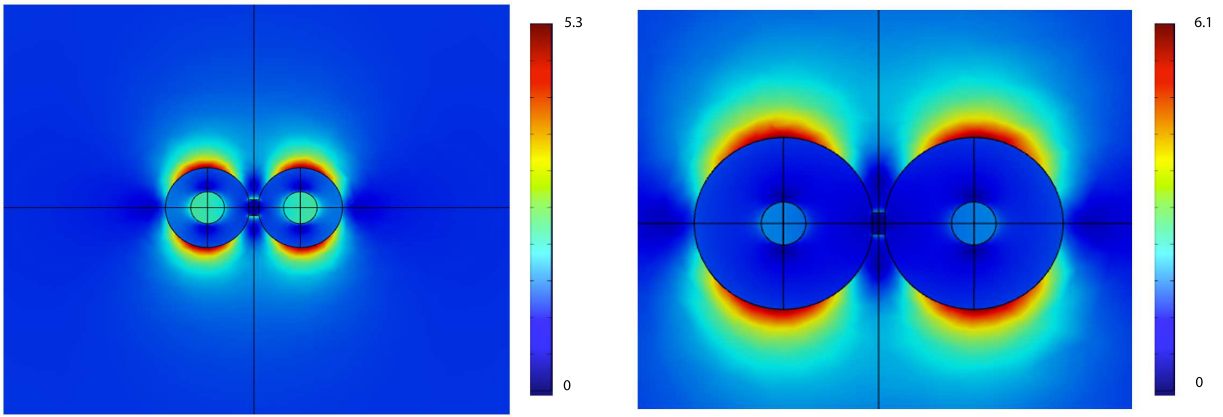}
\vskip 1.6cm
%\hfil
%\includegraphics[width=2.25in, angle=-0]{80.eps}
\caption{\sl Figure shows the variation of the UV absorption peak's intensity 
and the field distribution in the $XZ$ plane for the structure for    
outer radius of $50$~nm and $80$~nm.}
\end{center}   
\label{rlf8}  
\end{figure}   
%%%%%%%%%%%%%%%%%%%%%%%%%%%%%%%%%%%%%%%%%

\end{document}